\documentclass[iop]{emulateapj}
\usepackage{graphicx}
\usepackage[usenames,dvips]{color}
\usepackage{subfigure}

\newcommand{\lsim }{{\lower0.8ex\hbox{$\buildrel <\over\sim$}}}
\newcommand{\gsim }{{\lower0.8ex\hbox{$\buildrel >\over\sim$}}}

\def\apj{ ApJ}
\def\aap{ A\&A}

\def\aj{AJ}
\def\apjs{ApJ Supp}

\def\Chandra{${\it Chandra}$}

\def\simge{\mathrel{%
  \rlap{\raise 0.511ex \hbox{$>$}}{\lower 0.511ex \hbox{$\sim$}}}}
\def\simle{\mathrel{
  \rlap{\raise 0.511ex \hbox{$<$}}{\lower 0.511ex \hbox{$\sim$}}}}

\newcommand{\Msun}{\ifmmode {M_{\odot}}\else${M_{\odot}}$\fi}
\newcommand{\Lsun}{\ifmmode {L_{\odot}}\else${L_{\odot}}$\fi}
\newcommand{\Rsun}{\ifmmode {R_{\odot}}\else${R_{\odot}}$\fi}

\shorttitle{M13 Neutron Star Atmosphere}
\shortauthors{Catuneanu et al.}

\begin{document}
\title{Mass/Radius Constraints on the Quiescent Neutron Star in M13 Using Hydrogen and  Helium Atmospheres}

\author{Catuneanu, A.\altaffilmark{1,2}, Heinke, C.~O.\altaffilmark{1}, Sivakoff, G.~R.\altaffilmark{1}, Ho, W.~C.~G.\altaffilmark{3}, Servillat, M.\altaffilmark{4}}

\altaffiltext{1}{Dept. of Physics, University of Alberta, Room 238 CEB, Edmonton, AB T6G 2G7, Canada; heinke@ualberta.ca}
\altaffiltext{2}{Perimeter Institute, 31 Caroline St., N. Waterloo, ON, N2L 2Y5, Canada}
\altaffiltext{3}{School of Mathematics, University of Southampton, Southampton SO17 1BJ, UK}
\altaffiltext{4}{Laboratoire AIM, CEA Saclay, Bat. 709, 91191 Gif-sur-Yvette, France}

\begin{abstract}
The mass and radius of the neutron star (NS) in low-mass X-ray binaries can be
obtained by fitting the X-ray spectrum of the NS in quiescence,
and the mass and radius constrains the properties of dense matter in NS cores.
A critical ingredient for spectral fits is the composition of the NS
atmosphere: hydrogen atmospheres are assumed in most prior work, but helium
atmospheres are possible if the donor star is a helium white dwarf.  Here we
perform spectral fits to XMM, \Chandra, and ROSAT data of a quiescent NS in
the globular cluster M13.  This NS has the smallest inferred radius from
previous spectral fitting.  Assuming an atmosphere composed of hydrogen, we
find a significantly larger radius, more consistent with those from other
quiescent NSs.  With a helium atmosphere (an equally acceptable fit), we find even larger values for the radius.

\end{abstract}

\keywords{binaries : X-rays --- dense matter --- stars: neutron --- globular clusters: individual (NGC 6205)}

\maketitle

\section{Introduction}\label{s:intro}

Understanding the interiors of neutron stars (NSs) by measuring their masses and radii is a key goal of high-energy astrophysics \citep{Lattimer07}.  NS masses can be effectively  measured for radio pulsars \citep[e.g.][]{Antoniadis12} and some X-ray binaries \citep[e.g.][]{Rawls11} by radio timing and/or optical/IR radial velocities.  However, NS radii are much more complicated to measure. The radius measured at infinity is affected by the NS's gravitational redshift, as $R_{\infty}=R (1+z)=R/\sqrt{1-2 G M/(R c^2)}$ (where $R$ and $M$ refer to the NS values), and in more subtle ways by the surface gravity, magnetic fields, emission anisotropies, and composition.  Two key methods of constraining NS radii have involved fitting X-ray burst spectra \citep{Damen90}, and fitting X-ray spectra of quiescent low-mass X-ray binaries (qLMXBs) containing NSs \citep{Rutledge99}.  Recently, substantial work has been done using RXTE burst spectral measurements to constrain the mass and radius of NSs where the distance can be inferred, though the interpretation of these results has differed due to different choices of assumptions \citep{Boutloukos10,Ozel10,Steiner10,Suleimanov11b,Zamfir12,Galloway12}. 

 The qLMXBs typically show thermal blackbody-like radiation and/or a harder nonthermal component, often fit with a power-law of photon index 1 to 2 \citep{Campana98a}.  The nonthermal component is of uncertain origin, possibly due to continued accretion or a pulsar wind.  The thermal component is better understood, as emission from the NS surface, powered by some combination of continued accretion \citep{Zampieri95}, heat release from the crust deposited by the last outburst \citep{Rutledge02c,Degenaar11}, and heat deposited in the core (created by nuclear fusion in the deep crust during outbursts) that is now leaking out \citep{Brown98}. Fitting the thermal X-ray spectra of qLMXBs is conceptually simpler than fitting X-ray bursts, as accretion has either stopped or is at a very low level, allowing the stratification of the lightest element to the top of the atmosphere within 30 seconds \citep{Alcock80}. Calculations of nonmagnetic hydrogen atmosphere NS spectra generally reach very good agreement with each other and with observations \citep{Rajagopal96,Zavlin96,Heinke06a,Haakonsen12}, suggesting that X-ray spectra of qLMXBs can provide useful NS mass/radius constraints.  

However, 
qLMXB radius constraints depend strongly on the distance, 
 and the distances to most qLMXBs are poorly known.  One solution is to study qLMXBs in globular clusters, where the distances are typically known to $\sim5-10$\% precision \citep{Brown98, Rutledge02a}.  A number of qLMXBs have been studied in globular clusters \citep[see][for a review]{Guillot09a}, but only a few have sufficiently high-quality spectra to provide interesting constraints on the equation of state; these include X5 and X7 in 47 Tuc \citep{Heinke03a,Heinke06a}, and the qLMXBs in $\omega$ Cen \citep{Rutledge02a,Gendre03a, Webb07}, NGC 6397 \citep{Grindlay01b,Guillot11a},  M28 \citep{Becker03,Servillat12},  M13 \citep{Gendre03b,Webb07}, and NGC 6553 \citep{Guillot11b}.  Interestingly, two of these constraints (X7 in 47 Tuc, vs. M13) disagree with each other at the 99\% confidence level, motivating efforts to understand the discrepancy.  Uncertainties in the pileup correction used for 47 Tuc X7 \citep{Davis01} might drive the discrepancy, suggesting a deep observation of 47 Tuc with a smaller frame time to eliminate pileup.  Alternatively, the discrepancy may be due to differing atmospheres.

A large fraction of bright persistent or transient LMXBs in globular clusters have orbital periods less than one hour (so-called ``ultracompact'' systems, 5 of the 10 systems with known periods), requiring degenerate white dwarf companions; such systems can be easily created through dynamical interactions in globular clusters \citep{Deutsch00,Ivanova05}.  Ultracompact systems are likely to accrete material without hydrogen, since the donor star is (for the most typical evolutionary tracks) devoid of hydrogen, though spallation of infalling material to create hydrogen is possible \citep{Bildsten92}.  
Several characteristics of X-ray bursts differ between ultracompact vs.\ long-period LMXBs; the ratio of integrated persistent to burst fluence (much larger for ultracompact systems, indicating a lack of H), the existence of intermediate-long bursts at low mass transfer rates in ultracompacts (ignition of large He columns without H bursts), higher Eddington limits from ultracompacts than long-period sources (where distances are independently estimated from their globular cluster locations), and consistently short bursts at moderate mass transfer rates in ultracompacts (when mixed H/He ignition is expected if H is present) \citep{Cumming03,intZand05,Galloway08,Galloway10}.
Thus it seems likely that many globular cluster qLMXBs may have atmospheres composed of He, or C, depending on the nature of the donor star.

The possibility of differing atmospheric composition for globular cluster NS qLMXBs may explain the discrepancy between globular cluster qLMXB mass/radius constraints, since fits with He atmosphere models give larger radii than H atmosphere fits \citep{Ho09}.  We used the qLMXB in M28 as a first example of the differences in mass/radius constraints from the two models \citep{Servillat12}.  Here we consider the qLMXB with the smallest known radius constraint, the M13 qLMXB.  M13 has been studied by ROSAT's PSPC and HRI cameras, which detected an X-ray source (labeled Ga) in the core of the cluster \citep{Fox96,Verbunt01}, the target of this study.  \citet{Gendre03b} used XMM to identify another source, 15'' to the NW of Ga, which may contribute to the ROSAT PSPC and XMM spectra of Ga, and showed that Ga's spectrum was consistent with a hydrogen-atmosphere NS.  \citet{Webb07} then calculated detailed constraints on its mass and radius, along with XMM studies of two other NSs in globular clusters. 
\citet{Servillat11} used ground-based optical (Faulkes Telescope North), Chandra X-ray, and Hubble Space Telescope data to identify the source NW of Ga (their star 4, or X6; we use the latter name) as a cataclysmic variable (CV) experiencing a dwarf nova outburst.   Some additional results from the archival \Chandra\ observations of M13 (PI: Lewin) have been published \citep{Pooley06,Hui09}, but spectral analysis of \Chandra\ data on the NS qLMXB has not yet been published.

\section{Observations and Data Reduction}\label{s:obsdatared}

Our principal dataset is a pair of XMM-Newton observations obtained in Jan. 2002 using the EPIC cameras using the medium filters.  We also use a pair of \Chandra\  observations taken in March 2006 using the ACIS-S detector in FAINT mode, and a ROSAT observation taken in 1992 in pointing mode (see Table 1). 

\begin{deluxetable}{@{}ccccccc@{}}
\tablewidth{8cm}
\tiny
\tablecaption{\textbf{Observations}}
\tablehead{
\colhead{\textbf{Mission}} & \textbf{ObsID} &  \textbf{Date} &  \textbf{Instr.} &  \textbf{GTI} & \textbf{GTI2} & \textbf{Counts}\\
\colhead{} & & & & (s) & (s) & 
}
\startdata 
CXO & 7290 & 2006-03-09 & ACIS-S & 27894 & - & 300 \\
   CXO & 5436 & 2006-03-11 & ACIS-S & 26799 & - & 305 \\
   XMM & 85280301 & 2002-01-28 & MOS1 & 18044 & 14814 & 86 \\
  & & & MOS2 & 18051 & 15070 & 76 \\
   & & & PN & 14353 & 10338 & 283 \\
   XMM & 85280801 & 2002-01-30 & MOS1 & 16340 & 13537 & 126 \\
  & & & MOS2 & 16630 &  14004 & 51 \\
   & & & PN  & 12673 & 9025 & 428 \\
   ROSAT & 300181 & 1992-09 & PSPCB & 45872 & - & 452 \\
\enddata
\tablecomments{Observations of M13 used in this analysis, with GTI exposure times for XMM data quoted both with (GTI2) and without (GTI) aggressive background flare removal.  The number of counts in the extraction regions are given for our extractions without aggressive flare removal. }
\label{table:obsertable}
\end{deluxetable}

The \Chandra\ data were reduced using \Chandra\ Interactive Analysis of
Observations (CIAO) v.4.4\footnote{http://cxc.cfa.harvard.edu/ciao/} and \Chandra\ Calibration Database (CALDB) v4.4.8.  We reprocessed the data 
with the {\it chandra\_repro} reprocessing script to apply the
latest calibration updates and bad pixel files, and filtered the data to the energy range
$0.3-10.0$ keV.  The \Chandra\ data showed no strong 
background flaring, so we included all data. 

The position of the qLMXB in 
M13 (using {\tt wavdetect} on the combined \Chandra\ data) is 
R.A.$=16^h41^m43.77^s$, decl.$=+36^\circ27'57.64''$, consistent with the position given in \citet{Servillat11} for their \Chandra\ source X7.  The neighboring source, at R.A.=$16^h41^m42.47^s$, decl.$=+36^\circ28'07.29''$, is source X6 from \citet{Servillat11} (see Fig.\ 1), who showed it to be a CV exhibiting a dwarf nova eruption. 
Its X-ray luminosity is $L_X$(0.5-10)$=2.2^{+1.3}_{-1.1}\times10^{32}$ ergs/s, for a spectral shape consistent with an absorbed power-law of photon index $\Gamma=1.4\pm0.4$ (from simple spectral fits to the \Chandra\ spectra).  
The spectra of the qLMXB 
were extracted 
using circles with radii of $\sim2''$.  
The {\it specextract} script generated the corresponding
auxiliary response files (ARFs) and redistribution matrix files
(RMFs). 

The XMM-Newton data were reduced with the
Science Analysis System (SAS) v11.0.0. We repipelined the MOS data
using {\it emchain} and the PN data with {\it epchain} before applying the relevant 
filters (e.g. patterns $0-12$ for MOS, $0-4$ for pn). 
We also repipelined the MOS and PN data using the 2007 calibration data, as available to  \citet{Webb07}, for comparison to their analysis (see \S 3.2). 
Both XMM observations revealed signs of background flaring, affecting roughly 1/3 of the observations. Since the background flares are not extremely bright, we judge that spectra including all data attain a higher signal-to-noise ratio than if the flares were filtered out.  We do, however, filter out flares to match the GTIs of \citet{Webb07} when reproducing their analysis. In this case, we use filters of 5 and 10 counts per second for the two MOS datasets, and 30 and 50 counts per second for the pn data.

\begin{figure}
\includegraphics[width=0.5\textwidth]{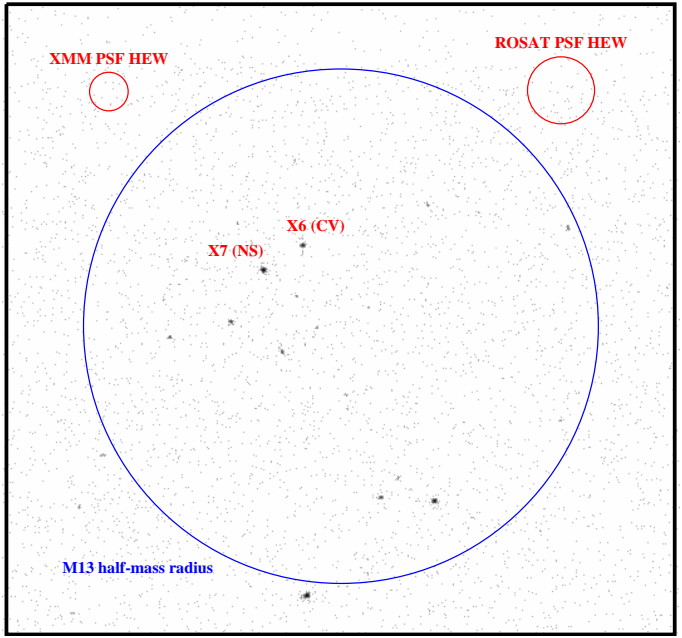}
\caption{Combined \Chandra\ 0.5-6 keV image of M13, showing the two relevant sources X7 (the NS qLMXB) and X6 (the nearby CV). The radius enclosing half of M13's mass (1.69') is indicated, as are circles roughly indicating the half-equivalent-width (HEW, enclosing 50\% of the energy) of the point-spread functions of the XMM-Newton pn (radius 7.6'') and ROSAT PSPC (radius 0.22') detectors. }
\end{figure}

Circular regions with
approximate radii of $9.5''$ (to exclude photons from the CV X6)  
were used to extract the
spectra of the NS qLMXB from the XMM observations, except for the analysis in \S 3.2, where we used a larger ($24''$) extraction circle, with a $13''$ circle around X6 excluded, to match the extraction region used in \citet{Webb07}.   ARFs
and RMFs were generated with the SAS tasks {\it arfgen} and {\it rmfgen}. 
We combined the two MOS spectra in each XMM observation using the {\it addspec} tool to achieve better statistics, except for the \S 3.2 analysis. Finally, all spectra were grouped to at least 20 counts per bin.

We used the {\it XSELECT} tool in  HEASOFT\footnote{http://heasarc.gsfc.nasa.gov/docs/software/lheasoft/} (v6.12) to reduce the ROSAT data. \footnote{http://heasarc.gsfc.nasa.gov/docs/rosat/ros\_xselect\_guide/xselect\_ftools.html}. The NS spectrum was extracted from a circular region of approximately $47''$ in radius and then grouped to twenty counts per bin with no filters applied, using the on-axis response matrices. 

\section{Data Analysis and Results} 

We first fit the hydrogen NS atmosphere to the combined XMM, \Chandra\ and ROSAT dataset, and calculate the constraints on the NS mass and radius.  We then attempt to reproduce the \citet{Webb07} result, using the data and calibrations available then.  Finally, we fit helium atmosphere models, and report the NS constraints in that case.  In all cases, we assume a distance to M13 of 7.7 kpc \citet{Harris96} (2003 revision), and a minimum $N_H$ of $1.1 \times 10^{20}$ cm$^{-2}$ \citep{Harris96}.

\subsection{NSATMOS Applied to Chandra, XMM and ROSAT Data}

We simultaneously fit the XMM-Newton, \Chandra, and ROSAT spectra to an absorbed  hydrogen atmosphere NS model, NSATMOS \citep{Heinke06a}.   The NS  mass was fixed to $1.4\Msun$ for our initial fitting (relaxed below).  We use a normalization constant to allow for the relative differences between each detector. 
 The cross-calibration work of \citet{Tsujimoto11} indicates that the MOS detectors give normalizations matching the average of several observatories, so we fix the MOS normalization to 1.0 and allow the pn normalization to float.  Fixing the pn normalization to 1.0 instead leads to slightly ($\sim$5\%) larger inferred NS radii, while fixing the \Chandra\ normalization leads to slightly smaller inferred radii.  

We also included a bremsstrahlung model when fitting the ROSAT data, as ROSAT's PSPC camera was unable to resolve the NS and X6 as individual point sources (see Fig.\ 1), following \citet{Webb07}. In the XMM and \Chandra\ data sets, these sources were resolved, with no evidence for a contribution by X6 to the NS spectrum (again agreeing with \citealt{Webb07}).  Since X6 provided few counts in the ROSAT data, and ROSAT has relatively poor spectral resolution (and no information above 2.5 keV), we fixed  $kT$ for the bremsstrahlung model of X6 to 4.5 keV in our fits, consistent with \citet{Webb07} and with simple fits to its \Chandra\ spectra.  Omitting this component had relatively small effects, but our ROSAT fits include this component as it should be present; its fitted 0.5-10 keV flux was $2\times10^{-13}$ ergs/cm$^2$/s.

 We also tried adding a power-law component to the model describing all data, as found for many qLMXBs, with a photon index fixed at either 1.5 or 2 \citep{Campana98a}.  Such a power-law did not significantly improve the fit, and the normalization was consistent with zero within 90\% confidence. To get the clearest constraints on a power-law component,  we fit the 0.3-8 unbinned \Chandra\ spectra with the C-statistic in XSPEC.  This finds the power-law normalization to be consistent with zero, with its upper limit to be 8\% of the total 0.5-10 keV flux, consistent with power-law upper limits from other globular cluster NS qLMXBs \citep{Heinke03d}.  We omit the power-law component from our fitting below.

\begin{deluxetable}{lc}
\tablewidth{8cm}
\tablecaption{\textbf{H Atm. Fits, CXO, XMM \& ROSAT}}
\tablehead{
\colhead{Parameter} & {Value} \\
}
\startdata
\multicolumn{2}{c}{\bf XMM Filtered for Flares}\\
\hline \\
Intrinsic $N_H$ & $0.0^{+0.8}_{-0.0} \times 10^{20} \ {\rm cm}^{-2}$ \\ 
NSATMOS Log$_{10}T$  & $5.99^{+0.05}_{-0.03}$ \\
NSATMOS $R$  & $11.7^{+1.9}_{-2.2} \ {\rm km}$ \\
Reduced $\chi^2$/dof  & 0.7551/73 \\
Null Hyp. Prob.  & 0.94 \\
\hline \\
\multicolumn{2}{c}{\bf Without Flare Filtering}\\
\hline \\
Intrinsic $N_H$ & $0.0^{+0.7}_{-0.0} \times 10^{20} \ {\rm cm}^{-2}$ \\ 
NSATMOS Log$_{10}T$  & $6.00^{+0.06}_{-0.04}$ \\
NSATMOS $R$  & $10.6^{+2.1}_{-2.2} \ {\rm km}$ \\
Reduced $\chi^2$/dof  & 0.8337/84 \\
Null Prob.  & 0.86 
\enddata
\tablecomments{\Chandra, ROSAT and XMM-Newton (flare-filtered and unfiltered) data fit to an NSATMOS hydrogen-atmosphere NS model. A bremsstrahlung model was added to the ROSAT spectrum to model X6's spectrum, included in the ROSAT extraction. The second fit corresponds to Fig. 3 for a NS mass fixed to $1.4 \Msun$. The $N_H$ quoted is any intrinsic $N_H$ in the binary, in addition to the (fixed) cluster value of $1.1\times10^{20}$ cm$^{-2}$.}
\label{table:table3}
\end{deluxetable} 
 
We have fit the model to both flare-filtered and unfiltered XMM data. While the flare-filtered data have less background contamination, they contain only 2/3 of the source photons, so the S/N ratio is similar.  Table 2 compiles results of fits to both cases, with the NS mass held fixed at $1.4\Msun$. We found no evidence for additional absorption in the binary.  We show a spectral fit to \Chandra, ROSAT and unfiltered XMM data in Fig.\ 2. 

\begin{figure}
\includegraphics[scale=0.33, angle=-90]{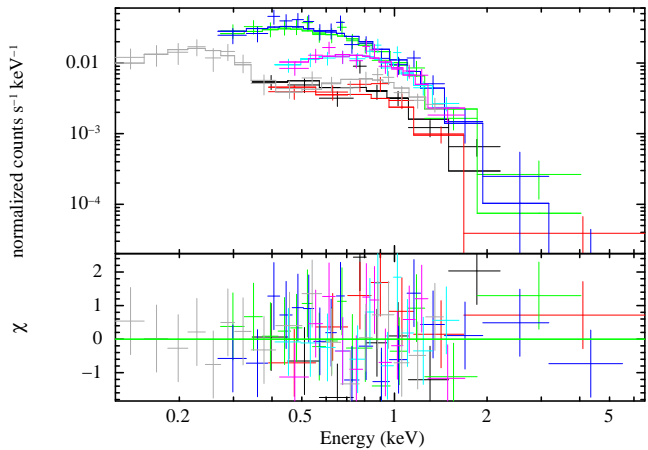}
\caption{Spectra of M13 qLMXB, including unfiltered XMM MOS (red and black) and pn (blue and green), \Chandra\ (light blue and magenta), and ROSAT (light gray) data, fit with the NSATMOS hydrogen-atmosphere model.  (The ROSAT model also contains a bremsstrahlung component, for the unresolved CV X6.) The lower portion of the diagram depicts $\chi^2$ residuals. }
\end{figure}

We then allow the mass of the NS to vary, finding a best fit with NS mass $1.5\Msun$ and radius 10.2 km (using the full XMM dataset). We show a contour plot of $\chi^2$, generated with the steppar command, over mass and radius in Fig.\ 3. The filtered XMM dataset gave similar contours, though the best-fit NS mass slid down the minimal $\chi^2$ valley to reach the lower bound of the NSATMOS model at $0.5 \Msun$.

\begin{figure}
\includegraphics[scale=0.37]{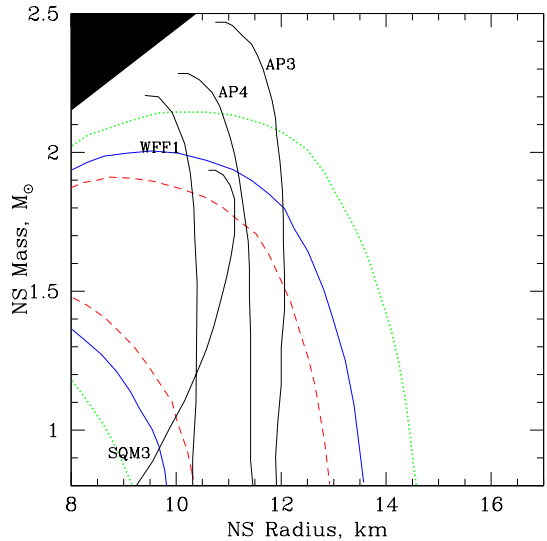}
\caption{Mass vs radius confidence contours for the non-flare-filtered fit to XMM, Chandra and ROSAT data with the NSATMOS hydrogen atmosphere. The solid (blue), dotted (green), and dashed (red) lines represent 90\%, 99\%, and 68\% confidence contours respectively. The upper left section (with $R<2.8 M$, \citealt{Lindblom84}) is shaded to indicate its inaccessibility for any NS, as any equation of state would require the sound speed to exceed $c$.}
\end{figure}

\subsection{NSATMOS Applied to XMM and ROSAT Data}

We noticed that our hydrogen-atmosphere model fit, above, did not reach similar results on the size of the M13 NS to the work of \citet{Webb07}, who require a relatively small NS  ($<$11 km at 90\% confidence for masses $\sim$1.4 \Msun).  We attempted to replicate their fit by reprocessing and extracting the XMM data as in Webb et al. (see \S 2 above).  For a NS mass fixed to 1.4 \Msun, we obtain the fit in Table 3. 

\begin{deluxetable}{lc}
\tablewidth{7cm}
\tablecaption{\textbf{H Atm. Fits, XMM \& ROSAT}}
\tablehead{
\colhead{Parameter} & {Value} \\
}
\startdata
Intrinsic $N_H$ & $0.0^{+0.6}_{-0.0} \times 10^{20} \ {\rm cm}^{-2}$ \\ 
NSATMOS Log$_{10}T$  & $6.04^{+0.06}_{-0.07}$ \\
NSATMOS $R$  & $9.5^{+3.0}_{-1.5*} \ {\rm km}$ \\
 Reduced $\chi^2$/dof  & 0.8595/68 \\
  Null Hyp. Prob. & 0.79 \\
\enddata
\tablecomments{ROSAT and flare-filtered XMM spectra of the M13 NS  fit by an NSATMOS model, with a bremsstrahlung component included to fit X6 in the ROSAT data.   This corresponds to Fig. 4 for a NS mass fixed to $1.4 \Msun$. Errors marked with * indicate the parameter hits a hard limit of the model.  The $N_H$ quoted is any intrinsic $N_H$ in the binary, in addition to the (fixed) cluster value of $1.1\times10^{20}$ cm$^{-2}$.}
\label{table:table2}
\end{deluxetable}

When allowing the mass to vary, we find a best-fit NS mass of $1.3\Msun$ and a $10.0$ km radius, not dissimilar to the best fit of Webb et al.\ (their Table 2). 
However, our contours (Fig.\ 4) are rather less constraining than those in Webb et al.'s Fig.\ 6.
Webb et al.'s quoted best fits lie very close to their upper 90\% confidence contour lines.  We do not see such behavior in any well-behaved chi-squared contour plots of the M13 qLMXB, or any other NSs in our experience.  The uncertainties on radius quoted in their Table 2 also seem unusually narrow.  

\begin{figure}
\includegraphics[scale=.37]{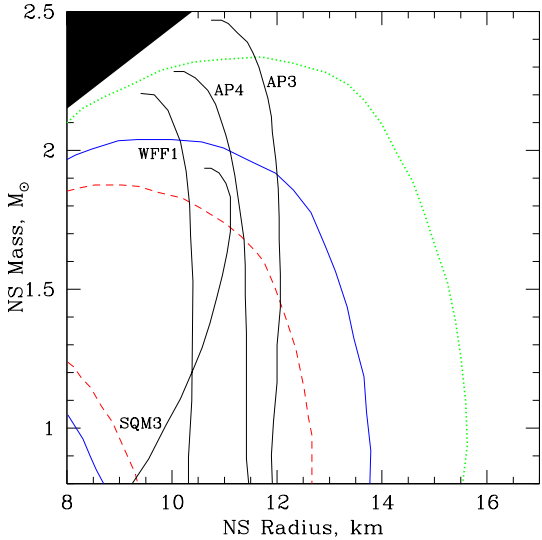}
\caption{Mass vs. radius contours from the NSATMOS hydrogen-atmosphere fit to XMM and ROSAT data processed following \citet{Webb07}. The solid (blue), dotted (green) and dashed (red) lines represent 90\%, 99\%, and 68\%  confidence contours respectively.  }
\end{figure}

\subsection{A Helium Atmosphere}

Finally, we consider the effects of a helium atmosphere on the inferred NS properties. 
We used the helium atmosphere model described in \citet{Ho09} to fit the data as described in \S 3.1.  We present fits with the NS mass fixed to 1.4 \Msun, using flare-filtered vs.\ unfiltered XMM data, in Table 4, and plot mass vs.\ radius contours in Fig.\ 5. 

\begin{deluxetable}{lc}
\tablewidth{8cm}
\tablecaption{\textbf{Helium Atm. Fits, CXO, XMM \& ROSAT}}
\tablehead{
\colhead{Parameter} & {Value} \\
}
\startdata
\multicolumn{2}{c}{XMM Filtered for Flares} \\
\hline \\
Intrinsic $N_H$ & $0.4^{+0.9}_{-0.4} \times 10^{20} \ {\rm cm}^{-2}$ \\
He NS Log$_{10}T$ &  $5.94^{+0.05}_{-0.04}$ \\
He NS $R$ &  $14.6^{+3.5}_{-3.1} \ {\rm km}$ \\
Reduced $\chi^2$/dof & 0.7753/73\\
Null Prob.  & 0.92 \\
\hline \\
\multicolumn{2}{c}{Without Flare Filtering} \\
\hline \\
Intrinsic $N_H$ & $0.2^{+0.8}_{-0.2} \times 10^{20} \ {\rm cm}^{-2}$ \\
He NS Log$_{10}T$  & $5.96^{+0.05}_{-0.05}$ \\
He NS $R$ &  $12.8^{+3.2}_{-1.2} \ {\rm km}$ \\
Reduced $\chi^2$/dof & 0.8341/84\\
Null Prob. &  0.86 \\
\enddata
\tablecomments{Fits to our helium atmosphere model.  The second fit corresponds to Fig. 5 for a NS mass fixed to 1.4 \Msun (see text).  The $N_H$ quoted is the intrinsic $N_H$ in the binary, added to the (fixed) cluster value of $1.1\times10^{20}$ cm$^{-2}$.} 
\label{table:table4}
\end{deluxetable} 

\begin{figure}
\includegraphics[scale=.37]{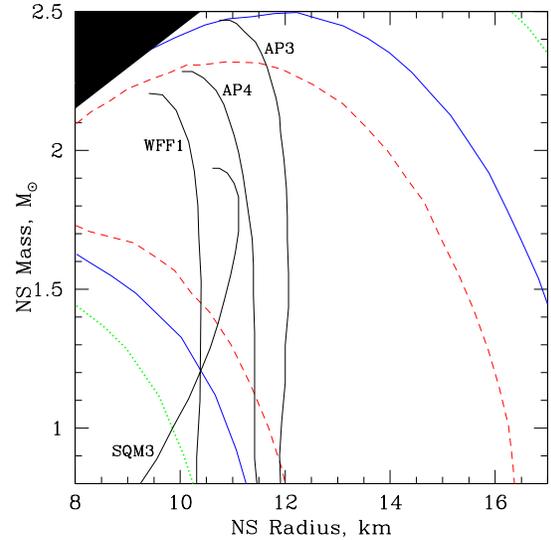}
\caption{Mass vs.\ radius contours when fitting the \Chandra, non-flare-filtered XMM, and ROSAT data with a helium atmosphere model.}
\end{figure}

Although the flare-filtered data have a slightly better fit to the hydrogen model ($\delta \chi^2=1.5$ for the same degrees of freedom), the helium model fit is still acceptable.  The fits to the unfiltered XMM data produce essentially indistinguishable $\chi^2$ values for helium or hydrogen fits. Thus, we cannot determine from the current data whether a hydrogen or helium atmosphere is the correct one, and thus which mass-radius contours are appropriate for the M13 qLMXB.

\section{Discussion}

Whether a NS qLMXB hosts a hydrogen or helium atmosphere affects the mass and radius constraints inferred from fitting its quiescent X-ray spectrum.  For the M13 qLMXB, for instance, the best-fit NS radius increases by $\sim$2 km with a helium atmosphere.  Although ultracompact LMXBs are known to be common in globular clusters, it is not certain whether ultracompact qLMXBs will display pure helium (or, for hybrid white dwarf donors, carbon; see \citealt{Nelemans10}) atmospheres, which introduces uncertainty into calculations of the radii of any NSs that may be in ultracompact qLMXBs.

There are several ways to distinguish ultracompact qLMXBs from normal qLMXBs.  Detection of hydrogen in the spectrum, or (narrow-band) photometry, of the companion can provide a clear distinction; this proves that the qLMXB in $\omega$ Cen has a hydrogen atmosphere \citep[e.g.][]{Haggard04}.  Measurement of the orbital period (e.g. through eclipses or pulsation timing in outburst) allows clear discrimination, as done for the eclipsers W37 and X5 in 47 Tuc \citep{Heinke05b,Heinke03a} and the millisecond pulsars SAX J1748.9-2021 and NGC 6440 X-2 in NGC 6440 \citep{Altamirano08,Altamirano10}.  The energetics and duration of X-ray bursts, when these can be clearly attributed to a particular qLMXB, may provide evidence for or against the presence of hydrogen; e.g., the surface of the Terzan 5 transient EXO 1745-248 is known to contain hydrogen \citep{Galloway08}.  For the M13 qLMXB, narrow-band photometry should be the first project.  Finding evidence of H-$\alpha$ emission from an  optical counterpart, using the \Chandra\ position and archival (or new, deeper) Hubble Space Telescope observations of M13, would prove that the NS photosphere is made of hydrogen.  

Another concern is the presence of cross-calibration differences between the various detectors used for this type of work; \Chandra's ACIS detector, and XMM's pn and MOS detectors.  \citet{Tsujimoto11} found that the XMM pn detector gives 1-8 keV fluxes 6.5\% lower than the MOS detectors in simultaneous observations, while \Chandra's ACIS-S3 detector averages 11.6\% higher fluxes.  Our analysis gave average XMM pn normalizations 5.5\% higher, and average \Chandra\ ACIS-S3 normalizations 12.5\% higher, than the MOS normalizations, consistent with other \Chandra/XMM cross-calibration in the 0.5-2 keV range \citep{Nevalainen10}.  Fixing the other instruments, instead, to have normalization equal to one would systematically increase the inferred NS radius by roughly 3 and 6\%, respectively.  It is not obvious which detector's calibration is more accurate, which suggests a $\sim$5\% systematic uncertainty in all such NS radius measurements.

Our result affects inferences of the equation of state of NSs, as the \citet{Webb07} analysis of the M13 NS gave some of the tightest constraints, consistent with the \citet{Ozel10} meta-analysis of X-ray burst results that preferred a relatively small NS radius under 10 km.  Our re-analysis does not support 
small radii for a hydrogen atmosphere fit.  For a helium atmosphere, the radius would be even larger, $>$10 km at 1 sigma for typical NS masses.  
Our results are consistent with the \citet{Steiner10} meta-analysis of NS bursters and qLMXBs, which preferred a NS radius between 11-12 km. \citealt{Steiner12} explicitly show that their results are robust against the removal of the \citet{Webb07} M13 qLMXB constraints.   Additional high-quality NS constraints would be very valuable in further constraining the NS equation of state, but should consider uncertainties in atmospheric composition and absolute flux calibration. 

\acknowledgements

We thank N. Webb and S. Morsink for discussions.  COH and GRS are supported by NSERC Discovery Grants, and COH also by an Ingenuity New Faculty Award.  WCGH appreciates the use of the computer facilities at KIPAC.  WCGH acknowledges support from STFC in the UK.

\bibliographystyle{apj}

\end{document}